\documentclass[a4paper,12pt]{iopart}
\usepackage{iopams}  
\usepackage{graphicx}

\newcommand\cL{\mathcal{L}}

\newcommand\Id{1}

\begin{document}

\title[Minimax mean estimator for the trine]{Minimax mean estimator for the trine}

\author{Hui Khoon NG$^{1,2}$, Kia Tan Benjamin PHUAH$^3$, and Berthold-Georg ENGLERT$^{1,3}$}
\address{$^1$ Centre for Quantum Technologies, National University of Singapore, 3 Science Drive 2, Singapore 117543, Singapore}
\address{$^2$ Applied Physics Lab, DSO National Laboratories, 20 Science Park Drive, Singapore 118230}
\address{$^3$ Department of Physics, National University of Singapore, 2 Science Drive 3, Singapore 117542, Singapore}
\ead{\mailto{cqtnhk@nus.edu.sg}, \mailto{ktphuah@gmail.com}, \mailto{cqtebg@nus.edu.sg}}

\begin{abstract}
We explore the question of state estimation for a qubit restricted to the $x$-$z$ plane of the Bloch sphere, with the trine measurement. 
In our earlier work [H.~K. Ng and B.-G. Englert, eprint arXiv:1202.5136[quant-ph] (2012)], similarities between quantum tomography and the tomography of a classical die motivated us to apply a simple modification of the classical estimator for use in the quantum problem. 
This worked very well.
In this article, we adapt a different aspect of the classical estimator to the quantum problem.
In particular, we investigate the mean estimator, where the mean is taken with a weight function identical to that in the classical estimator but now with quantum constraints imposed.
Among such mean estimators, we choose an optimal one with the smallest worst-case error---the minimax mean estimator---and compare its performance with that of other estimators.
Despite the natural generalization of the classical approach, this minimax mean estimator does not work as well as one might expect from the analogous performance in the classical problem.
While it outperforms the often-used maximum-likelihood estimator in having a smaller worst-case error, the advantage is not significant enough to justify the more complicated procedure required to construct it. 
The much simpler adapted estimator introduced in our earlier work is still more effective.
Our previous work emphasized the similarities between classical and quantum state estimation; in contrast, this paper highlights how intuition gained from classical problems can sometimes fail in the quantum arena.
\end{abstract}



\submitto{\NJP}

\maketitle

\section{Introduction}

Tomography is the process of characterizing a physical system.
In the simplest scenario, it estimates a single parameter of concern, for example, the transmission probability of a beam-splitter used in an optics experiment.
In the most general case, tomography involves estimating the state of the system, which provides the complete description of all properties of the system.
Here, we focus on the latter case of state tomography.

Tomography is an old subject, well explored first in the classical context
(see, for example, Ref.~\cite{LehmannCasella}), later in quantum scenarios
(see, for example, the classic textbook by Helstrom \cite{HelstromBook}),
and is still a very active area of research. For a reasonably recent and
comprehensive review of developments in quantum tomography, we point the
reader to the collection of articles found in Ref.~\cite{ParisBook}. Here,
we focus only on a general overview of the topic, and provide just
sufficient background for the reader to understand the context of our
current discussion.

Tomography involves two steps: 
first, the measurement of many identical copies of the system, which requires a choice of measurement; 
second, the estimation of the quantity of interest---be it a single parameter, or the full state---from the gathered measurement data, which requires a choice of estimator.
The choice of measurement can vary from single-copy measurements where one measures one copy at a time, to a  joint measurement implemented on all available copies of the system at once.
We focus here on single-copy measurements.
In particular, we discuss the simplest case of having the same measurement on every copy of the state, as opposed to adaptive strategies where the measurement to be performed on subsequent copies is modified according to the data collected from previous copies.

The choice of estimators---mathematically describable as maps from the set of possible data to the set of possible states---is equally varied. 
Estimation theory, as discussed by statisticians, explore estimators from the often-used maximum-likelihood (ML) estimator, to classes of estimators like Bayes estimators, minimax estimators, etc., each motivated by its own philosophy of inference from the data. 
These are all instances of \emph{point} estimators where one provides a single state (as opposed to a set of states for \emph{region} estimators) as the result of the tomography. 

Estimation theory, originally invented in the context of classical problems, is also applicable to the estimation of quantum states. 
By \emph{classical}, we simply mean that there exists a preferred basis for the system, and all states are described by \emph{probabilistic} mixtures of basis states. 
In contrast, quantum systems do not possess such a preferred basis, and describing quantum states requires not just probabilities, but probability amplitudes.
This difference between quantum and classical systems complicates the issue of state estimation.
Taking a frequentist's perspective, the relative frequencies of the measurement outcomes computed from a given set of data should approximate the probability of getting each outcome for the input state.
A good guess for the state will thus be one with outcome probabilities equal or close to the obtained relative frequencies.
For classical systems, every probability distribution corresponds to a physical state of the system.
Not so for quantum systems.
Outcome probabilities for a physical state satisfy a set of constraints dictated by quantum mechanics, but relative frequencies are unconstrained.
In fact, situations exist where violation of the constraints by treating relative frequencies as probabilities is generic rather than unusual.
Naively equating outcome probabilities to relative frequencies and using this to reconstruct the quantum state can thus result in an unphysical estimator.

Nevertheless, we can usefully think of quantum state estimation as classical state estimation with constraints imposed on the probability distributions describing the states. 
One needs to then invent methods of modifying estimators from classical estimation theory to enforce these constraints. 
In \cite{Ng12}, we did this by an ad-hoc ``minimal correction", by admixing the completely mixed state to the classical estimator, by an amount chosen in a minimax way, to render the resulting estimator---which we refer to as the \emph{corrected minimax estimator}---physical. 
More generally, one can modify the estimators by adopting the same inference philosophy as in the analogous classical problem, but now incorporating the physicality constraints required by quantum mechanics.
For example, ML methods prescribe a \emph{constrained} maximization of the likelihood function over the set of physical quantum states (see, for example, \cite{MaxLikRev}), as opposed to reporting the (unconstrained) maximum of the likelihood function as the estimator, which may happen to be outside of the set of physical states.

In this work, we adopt the minimax philosophy that leads to a good estimator for the classical die problem, and examine the analogous estimator for the problem of tomography of a qubit. 
To restrict to the simplest case with quantum features, we consider tomography of a qubit state, with the promise that the state of the system lies solely in the two-dimensional $x$-$z$ plane of the Bloch sphere.
Albeit a rather artificial promise, one can view this as tomography of a qubit system where we are interested only in the information restricted to the $x$-$z$ plane.
For example, this is of practical relevance to the four-state BB84 quantum key distribution scheme \cite{BB84}, which uses only a pair of conjugate bases of states lying entirely in the $x$-$z$ plane, and hence only information pertaining to that plane is of relevance.
We make use of the \emph{trine} measurement, with outcomes as subnormalized projectors onto the trine states---three pure states symmetrically arranged in the $x$-$z$ plane of the Bloch sphere.
The trine measurement is informationally complete for the $x$-$z$ plane of the qubit Bloch sphere.

In \cite{Ng12}, we showed how to make use of the similarities of between classical and quantum state tomography to construct simple estimators for the quantum problem that perform well and inherit desirable properties of the classical estimator. 
In contrast, here we show that, despite the similarities, applying the same philosophy that worked well in the classical die problem to our trine problem, the seemingly minor and rather natural modification of the classical estimator turns out to work poorly in the quantum case. 
Although this modified estimator still gives better performance (as quantified by the mean squared error) than the ML estimator, the additional computation effort needed to gain the small advantage seem hardly worth the trouble. 
This is particularly so given that our simple ad hoc procedure in \cite{Ng12} gives significantly better results. 
While \cite{Ng12} highlighted the similarities between classical and quantum state estimation, here we show how simple quantum constraints on the probabilities can greatly complicate matters and lead to results noticeably divergent from the classical case.

In the next section, we begin with the problem of state estimation for a classical $K$-sided die. 
We review the notion of a mean estimator, and explain the minimax approach to finding the optimal estimator. 
Using the mean estimator motivated by the classical problem, in Section \ref{sec:quantum}, we examine the problem of estimating a qubit state with the trine measurement. 
Again, we follow a minimax approach in choosing the optimal mean estimator, and compare its performance with that of the ML estimator and the corrected minimax estimator. We conclude in Section \ref{sec:conc}.

\section{The classical $K$-sided die}
We begin our discussion by considering a classical $K$-sided die, for which we are interested in finding out the different weighting of the faces of the die. This problem is well-known in the classical literature; our review of it here serves to define the notation and also motivate the ideas to be used later when studying the quantum problem.

\subsection{The measurement: a die toss}
We toss the die---a measurement---and ask which face of the die turns up.
The set of outcome probabilities provides a complete description of the die.
To phrase this in a more formal language suitable for discussing tomography, to each face of the die, we ascribe a pure state $|k\rangle$ such that $\langle k|l\rangle=\delta_{kl}$, for $k,l=1,2,\ldots, K$. 
The different faces of the die thus correspond to orthonormal states, and $\{|k\rangle\}_{k=1}^K$ is the preferred basis for the die. The die is described by a probability distribution $\{p_k\}_{k=1}^K$, where $\sum_{k=1}^K p_k=1$ and $p_k\geq 0$ for all $k$.
Each $p_k$ describes the probability that face $k$ turns up when the die is tossed.
We can also write the state of the die as a positive semi-definite operator $\rho$ with unit trace given by
\begin{equation}\label{eq:dieState}
\rho=\sum_{k=1}^K |k\rangle p_k\langle k|.
\end{equation}

A single toss of the die can be described by a \emph{probability operator measurement} (POM), with outcomes $\Pi_k\equiv |k\rangle\langle k|$, $k=1,2,\ldots, K$.
$\Pi_k$ is associated with the outcome that the $k$th face of the die turns up in a toss, and Born's rule gives $p_k\equiv \tr(\rho\Pi_k)$ as the outcome probabilities for the state $\rho$.
A die toss is an instance of a symmetric measurement, or an ``S-POM" (see Appendix A of \cite{Ng12} for more details on S-POMs).
Every S-POM has, apart from the outcome operators $\{\Pi_k\}_{k=1}^K$, a set of hermitian, trace-1 operators $\{\Lambda_k\}_{k=1}^K$ with the defining property that $\tr\{\Pi_k\Lambda_l\}=\delta_{kl}$. 
This allows the expansion of the part of the state measured by the S-POM as
\begin{equation}\label{eq:LambdaState}
\rho=\sum_{k=1}^Kp_k\Lambda_k.
\end{equation}
As detailed in \cite{Ng12}, the $\Lambda_k$ operators can be explicitly constructed given the $\Pi_k$ operators for an S-POM.
For the case of the classical die toss, $\Lambda_k=\Pi_k=|k\rangle\langle k|$.
Comparing with \eref{eq:dieState}, we see that \emph{every} state of the die can be written as in \eref{eq:LambdaState}.
Thus, the die toss is also \emph{informationally complete} (IC), in that it measures all aspects of the information pertaining to the die. 
The die toss is thus an example of a symmetric, informationally complete POM, or SIC-POM for short.

Repeated tosses of the die can be thought of as repeated measurements on multiple, identical copies of the die. 
Measurement on $N$ copies yields data $D_N\equiv \{c_1,c_2,\ldots, c_N\}$, where $c_i=1,2,\ldots,K$ indicates the outcome obtained in the measurement of the $i$th copy. The data can be summarized as $D_N\sim\{n_1,n_2,\ldots n_K\}$, where $n_k$ is the number of ``clicks" in the $k$th detector, indicating the number of times outcome $k$ was obtained. Note that $\sum_{k=1}^Kn_k=N$.

\subsection{The mean estimator}

Associated with the data $D_N\sim\{n_1,n_2,\ldots,n_K\}$ is the likelihood function 
\begin{equation}
\cL(D_N|\rho)\equiv \prod_{k=1}^K p_k^{n_k},
\end{equation}
which is the probability of obtaining data $D_N$ given a state $\rho$. 
ML methods suggest one to use as the estimator, the state that gives the largest likelihood for the data $D_N$. For the classical die, this yields the estimator
\begin{equation}
\hat\rho_\mathrm{ML}(D_N)\equiv\sum_k (\hat p_k)_\mathrm{ML}\Lambda_k=\sum_k (\hat p_k)_\mathrm{ML}|k\rangle\langle k|, ~~\textrm{with }(\hat p_k)_\mathrm{ML}\equiv \frac{n_k}{N}.
\end{equation}

Alternatively, one can view the likelihood as a \emph{weight} over states, and choose as our estimator, the weighted average over all states. 
This gives the \emph{mean estimator}, well known in classical state estimation,
\begin{equation}\label{eq:meanEst}
\hat \rho_\mathrm{ME}(D_N)\equiv \frac{\int \rmd\phi(\rho)\cL(D_N|\rho)\rho}{\int \rmd\phi(\rho)\cL(D_N|\rho)},
\end{equation}
where $\rmd\phi$ is an integration measure that tells us how to sum over states. $\rmd\phi$ is non-negative on physically permissible $\rho$, and zero elsewhere. 
A Bayesian approach to state estimation will set $\rmd\phi$ as the prior distribution, encompassing all prior information one has about the system to be characterized. 
The mean estimator in this case will then simply be the average state of the posterior distribution $\rmd\phi(\rho)\cL(D_N|\rho)$ (see, for example, Ref.~\cite{BlumeKohout10} for a recent discussion of the Bayesian mean approach to quantum tomography).
More generally, $\rmd\phi$ is can be thought of as a functional parameter that characterizes the class of all mean estimators.

Parameterizing $\rho$ by the outcome probabilities $p_k$s via \eref{eq:LambdaState}, $\int\rmd\phi$ can be written as
\begin{equation}\label{eq:intMeas}
\int\rmd\phi(p)(\ldots)=\int (\rmd p)\chi(p)f(p)(\ldots),
\end{equation}
where $p$ denotes the list of probabilities $\{p_1,p_2,\ldots,p_K\}$, $(\rmd p)\equiv \rmd p_1\ldots\rmd p_K$, $\chi(p)$ is a characteristic function that accounts for physicality constraints on $p$, and $f(p)$ is a non-negative weight function that can be adjusted to optimize the performance of the estimator with respect to a desired figure-of-merit.
The mean estimator can thus be written as 
\begin{equation}
\fl\qquad\hat\rho_\mathrm{ME}(D_N)\equiv\sum_k (\hat p_k)_\mathrm{ME}\Lambda_k,\quad\textrm{with }(\hat p_k)_\mathrm{ME}\equiv \frac{\int_0^\infty (\rmd p)\chi(p)f(p)\cL(D_N|p)~p_k}{\int_0^\infty (\rmd p)\chi(p)f(p)\cL(D_N|p)},
\end{equation}
where the lower integration limit of 0 ensures $p_k\geq 0$ for all $k$.
For the classical die,
\begin{equation}
\chi(p)=\delta{\bigg(1-\sum_k p_k\bigg)},
\end{equation}
where the delta function enforces that the probabilities $p_k$ sum to 1. 
This restricts the integration to over physical states of the classical die only.

The invariance of the physical properties of the die under the interchange of the (arbitrarily assigned) labels $k$ for the faces suggests consideration of an $f(p)$ that is unchanged under a permutation of the labels.
The form of the likelihood function further hints at an $f(p)$ given by
\begin{equation}\label{eq:f}
f(p)=\left(\prod_{k=1}^K p_k\right)^{\beta-1}.
\end{equation}
Existence of the integrals defining the mean estimator for the classical die requires $\beta>0$.
For this choice of $f(p)$, it is convenient to define moments
\begin{eqnarray}
M_\beta(n_1,n_2,n_3)&\equiv 
2\int_0^\infty (\rmd p)~\delta{\Big(1-\sum_kp_k\Big)}~\prod_kp_k^{n_k+\beta-1},\label{eq:Mdie}
\end{eqnarray}
where the normalization factor of 2 is chosen for convenience of the quantum problem to be discussed later.
The mean estimator, for a given value of $\beta$, can then be written as a ratio of two moments,
\begin{equation}\label{eq:pkMERatio}
(\hat p_k)_\mathrm{ME}^{(\beta)}=\frac{M_\beta(n_1,\ldots,n_{k-1},n_k+1,n_{k+1},\ldots,n_K)}{M_\beta(n_1,\ldots,n_{k-1},n_k,n_{k+1},\ldots,n_K)}.
\end{equation}
The moments for the classical die can be evaluated explicitly,
\begin{equation}\label{eq:dieM}
M_\beta(n_1,n_2,\ldots,n_K)=\frac{2~\Gamma(n_1+\beta)~\Gamma(n_2+\beta)~\ldots~\Gamma(n_K+\beta)}{\Gamma(N+K(\beta-1))},
\end{equation}
where $\Gamma(z)$ is the familiar Gamma function.
Since $\Gamma(z+1)=z\Gamma(z)$, we thus have
\begin{equation}\label{eq:addbeta}
(\hat p_k)_\mathrm{ME}^{(\beta)}=\frac{n_k+\beta}{N+K\beta},
\end{equation}
which can be rewritten as
\begin{equation}\label{eq:pME}
(\hat p_k)_\mathrm{ME}^{(\beta)}=\frac{1}{K}a^{(\beta)}_N+\frac{n_k}{N}~b^{(\beta)}_N,~
\textrm{ with }a^{(\beta)}_N\equiv \frac{1}{1+\frac{N}{\beta K}}~\textrm{ and }b^{(\beta)}_N\equiv \frac{1}{1+\frac{\beta K}{N}}.
\end{equation}

\subsection{The minimax estimator}
How should we choose the value of the parameter $\beta$?
We make use of a minimax approach: 
$\beta$ is chosen to minimize the worst-case (over all physical states) mean squared error (MSE), defined for state $\rho$ with outcome probabilities $p$ and estimator $\hat\rho$ with outcome probabilities $\hat p$ as
\begin{equation}
\textrm{MSE}(\rho,\hat \rho)\equiv \sum_{D_N}\cL(D_N|\rho)\sum_k{\left[p_k-\hat p_k(D_N)\right]}^2.
\end{equation}
While an arbitrary choice for quantifying the estimation error for using estimator $\hat\rho$, the MSE is used here since it is particularly amenable to analytical manipulations. 
It is an often-used measure of estimation error in classical problems. More generally, tomography with a SIC-POM gives an MSE that is equal, apart from an overall irrelevant constant factor, to the mean squared Hilbert-Schmidt distance between $\rho$ and $\hat\rho$: $\textrm{MSE}(\rho,\hat\rho)\propto \sum_{D_N}\cL(D_N|\rho)\tr{\{(\rho-\hat\rho)^2\}}$. This relation holds even for quantum tomography.
The average over measurement data is natural from the point of view of choosing a single estimation strategy that works for many runs of tomography of the same state, each of which can yield a different $D_N$.

For the mean estimator for the classical die given in \eref{eq:pME}, the MSE is given by
\begin{eqnarray}
\textrm{MSE}{\Big(\rho,\hat\rho_\mathrm{ME}^{(\beta)}\Big)}
&=\frac{1}{(N+K\beta)^2}{\left[(\beta^2K^2-N)p^2+(N-\beta^2K)\right]},
\end{eqnarray}
where $p^2\equiv \sum_k p_k^2$.
Noting that $\frac{1}{K}\leq p^2\leq 1$ for any state of the classical die, we can consider the maximum (over all $\rho$) of the $\textrm{MSE}$ for three different cases: (i) $\beta^2 K^2>N$, (ii) $\beta^2K^2=N$, and (iii) $\beta^2K^2<N$. 
The minimum (over $\beta$) of the maximum MSE is attained for case (ii), where 
\begin{equation}\label{eq:betaDie}
\beta=\frac{\sqrt N}{K}, 
\end{equation}
yielding the \emph{minimax mean estimator} for the classical die,
\begin{equation}\label{eq:dieEst}
 (\hat p_k)_\mathrm{MM}=\frac{1}{K}a_N+\frac{n_k}{N}b_N,~\textrm{ with }
a_N\equiv \frac{1}{1+\sqrt N}\textrm{ and } b_N\equiv \frac{1}{1+1/\sqrt N}.
\end{equation}

Actually, the estimator in \eref{eq:dieEst} is minimax not just over the class of mean estimators parameterized by $\beta$, but is minimax over all estimators for the classical die problem. 
To see this, we observe that $\beta=\sqrt{N}/K$ results in an MSE that is constant over all states.
It is known from estimation theory that a mean estimator with constant MSE is minimax, that is, it has the smallest worst-case MSE, over all estimators (see, for example, \cite{LehmannCasella}; or see \cite{Ng12} for a self-contained proof of the fact).
This thus provides an objective justification for choosing a weight function $f(p)$ of the form \eref{eq:f}.

\section{The qubit confined to a plane}\label{sec:quantum}
We now turn to the quantum problem of a qubit confined to the $x$-$z$ plane of the Bloch sphere and explain how to adopt the same philosophy that led to the minimax estimator for the classical die in the quantum problem.

\subsection{The trine measurement and physicality constraints}
As stated in the introduction, the trine POM is an S-POM with three POM outcomes built from three pure states symmetrically arranged in the $x$-$z$ plane of the Bloch sphere, subtending angles of $\frac{2\pi}{3}$ between pairs of states. 
These states are collectively known as the \emph{trine states},
\begin{equation}
|\psi_k\rangle\langle\psi_k|=\frac{1}{2}\left(1+\sigma_z\cos\phi_k+\sigma_x\sin\phi_k\right),~\textrm{with }\phi_k\equiv\phi_0+(k-1)\frac{2\pi}{3},
\end{equation}
for $k=1,2,3$, where $\sigma_i$s are the usual Pauli operators for describing two-dimensional systems.
Here, $\phi_0$ is a fixed angle that determines the orientation of the trine states in the $x$-$z$ plane. 
The trine states are linearly independent and complete since
\begin{eqnarray}\label{eq:trinestate1}
|\langle\psi_k|\psi_l\rangle|^2=\frac{3}{4}\delta_{kl}+\frac{1}{4}\quad\textrm{and}\quad
\frac{2}{3}\sum_{k=1}^3|\psi_k\rangle\langle\psi_k|=\Id.
\end{eqnarray}
The outcomes of the trine POM are subnormalized projectors onto the trine states,
\begin{equation}
\Pi_k\equiv|\psi_k\rangle\frac{2}{3}\langle\psi_k|,\quad k=1,2,3.
\end{equation}
As is required for a physical POM, \eref{eq:trinestate1} ensures $\sum_{k=1}^3\Pi_k=1$.
Outcome probabilities for state $\rho$ using the trine POM are given by
\begin{equation}\label{eq:pkrphi}
p_k=\tr(\rho\Pi_k)=\frac{1}{3}[1+r\cos(\phi-\phi_k)],\quad k=1,2,3,
\end{equation}
where we have used the fact that every physical state lying in the $x$-$z$ plane of the Bloch sphere can be described in polar coordinates as 
\begin{equation}\label{eq:quantumState}
\rho=\frac{1}{2}{\left[1+r(\sigma_z\cos\phi+\sigma_x\sin\phi)\right]},\quad\textrm{with }0\leq r\leq 1\textrm{ and }0\leq \phi<2\pi.
\end{equation}
The $\Lambda_k$ operators for the trine measurement are
\begin{equation}\label{eq:Lambda}
\Lambda_k\equiv |\psi_k\rangle 2 \langle\psi_k|-\frac{1}{2}=\frac{1}{2}+\sigma_z\cos\phi_k+\sigma_x\sin\phi_k,\quad k=1,2,3,
\end{equation}
and a state with trine outcome probabilities $p_k$ can be written as $\rho=\sum_k p_k\Lambda_k$, as previously explained.
Every qubit state in the $x$-$z$ plane of the Bloch sphere can be described this way, with a unique set of outcome probabilities, demonstrating that the trine measurement is informationally complete for this restricted set of qubit states.

\begin{figure}
\begin{center}
\includegraphics[width=0.45\textwidth]{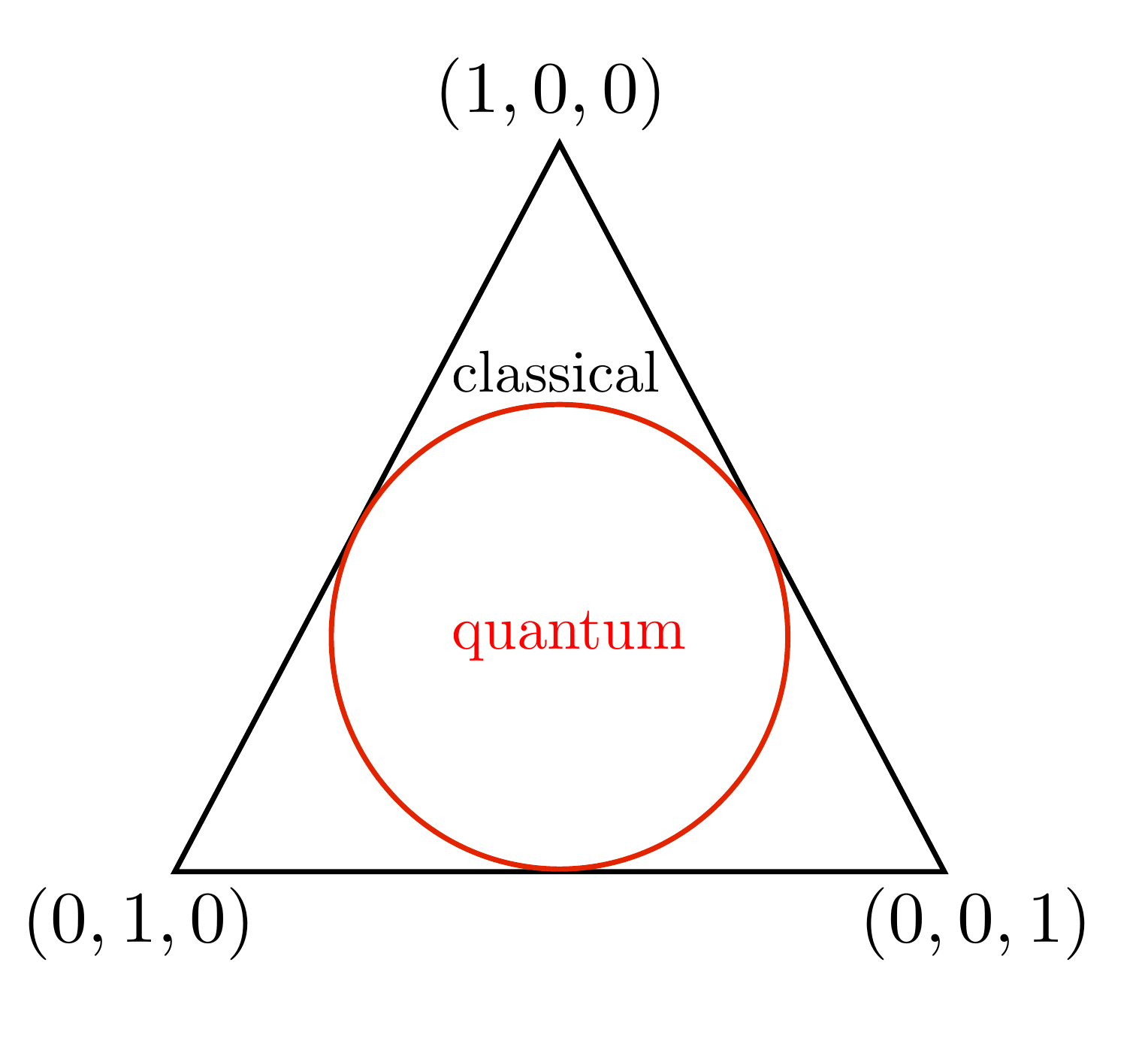}
\caption{\label{fig:triangleCircle} The triangle contains all points $(p_1,p_2,p_3)$ such that $\sum_kp_k=1$ and $p_k\geq 0$ for all $k$. Every point in (or on) the triangle corresponds to a physically permissible state of the classical 3-sided die. The disk contains all points in the triangle that also satisfy the quantum constraint of $\sum_k p_k^2\leq \frac{1}{2}$. Every point in the disk represents a physically permissible state of the qubit, while points in the triangle outside the disk are unphysical.}
\end{center}
\end{figure}

From \eref{eq:pkrphi}, the trine outcome probabilities satisfy
\begin{equation}\label{eq:trineConstr}
\frac{1}{3}\leq p^2\equiv \sum_{k=1}^3 p_k^2=\frac{1}{3}{\left(1+\frac{1}{2}r^2\right)}\leq \frac{1}{2},
\end{equation}
for all physical qubit states.
Here, we have used the identities $\sum_k\sin\phi_k\cos\phi_k=0$ and $\sum_k(\cos\phi_k)^2=\sum_k(\sin\phi_k)^2=\frac{3}{2}$.
Notice the difference here from the classical die problem. 
For a 3-sided classical die, the outcome probabilities satisfy no additional constraint apart from those that ensure they form a probability distribution (that is, $\sum_kp_k=1$ and $p_k\geq 0$ for all $k$, which also guarantee $p^2\geq \frac{1}{3}$), and $p^2$ can be as large as 1.
We can visualize the physical states of the classical 3-sided die as points on an equilateral triangle (the planar region defined by $\sum_kp_k=1$ and $p_k\geq 0$ for all $k$ in the $p_1$-$p_2$-$p_3$ space), with vertices corresponding to the states with outcome probabilities $(p_1,p_2,p_3)=(1,0,0), (0,1,0)$ and $(0,0,1)$ (see Figure \ref{fig:triangleCircle}).
Physical qubit states, with associated trine outcome probabilities, however, do not occupy the entire triangle. 
For example, the point $(1,0,0)$ does not correspond to a qubit state, since the outcome probabilities violate constraint \eref{eq:trineConstr}.
Instead, physical qubit states reside on the disk inscribed within the classical equilateral triangle (the intersection of the equilateral triangle with the ball of radius $\frac{1}{\sqrt{2}}$ in the $p_1$-$p_2$-$p_3$ space). 
Points in the triangle outside of the disk correspond to states with at least one negative eigenvalue, and are hence not permissible qubit states.

An estimator for the classical 3-sided die problem will report a point in the triangle. 
Estimators for the qubit with the trine measurement will, however, need to land only inside the disk.
The ML procedure instructs us to look for the maximum of the likelihood function, for given data, constrained to the disk. 
Our procedure in \cite{Ng12} tells us to start with the estimator for the 3-sided die problem, and if it lies outside of the disk, to``pull it in" towards the centre until it lies on the boundary of the disk (or better yet, just inside the boundary---see \cite{Ng12} for more details).
Below, we examine yet another approach to dealing with the constraint to the disk.

\subsection{The mean estimator for the trine}
Recall that the integration measure in the mean estimator (see \eref{eq:intMeas}) includes the characteristic function $\chi(p)$, which accounts for all physicality constraints. 
A natural way, then, to impose the constraint of $p^2\leq \frac{1}{2}$ for the trine in the mean estimator, is to include it explicitly in $\chi(p)$,
\begin{equation}
\chi(p)\equiv \delta{\left(1-\sum_{k=1}^3p_k\right)}~\eta{\left(\frac{1}{2}-\sum_{k=1}^3p_k^2\right)}.
\end{equation}
Here, $\eta(~)$ is Heaviside's step function: $\eta(x)=0$ for $x<0$ and $\eta(x)=1$ for $x>0$.
As in the classical die problem, the delta function ensures $\sum_kp_k=1$, while $p_k\geq 0$ is enforced by the lower limit of integration in the mean estimator. 
Any mean estimator, with the above choice of $\chi(p)$, will automatically  be a convex sum (integral) only of physical qubit states (confined to the $x$-$z$ plane of the Bloch sphere), and will hence be itself a physical state, or equivalently, will lie in the disk.

We are still left with the choice of the weight function $f(p)$. 
Apart from the additional constraint on $p^2$, the trine problem has identical symmetries as the 3-sided die problem.
This suggests consideration of the same $f(p)$ function that worked well for the classical die (see \eref{eq:f}),
\begin{equation}
f(p)={\left(\prod_{k=1}^K p_k\right)}^{\beta-1}.
\end{equation}
$\beta$ can be any real number as long as all the integrals in the mean estimator (for all data) exist. 
For the classical die, we needed $\beta>0$; 
for the trine, the additional step function in $\chi$ extends the range to $\beta>-\frac{1}{2}$.

As in the case of the classical die, we can define moments for the trine problem,
\begin{equation}\label{eq:Mtrine}
M_\beta(n_1,n_2,n_3)=2\int_0^\infty (\rmd p) ~\delta{\left(1-\sum_lp_l\right)}~\eta{\left(\frac{1}{2}-p^2\right)}\prod_{k=1}^3p_k^{n_k+\beta-1}.
\end{equation}
The normalization factor of 2 is chosen such that $M_1(0,0,0)=1$.
The trine moments look identical to those for the classical die (Eq.~\eref{eq:Mdie}), except for the additional step function, which contains the only visible quantum-mechanical feature of the problem.

The seemingly harmless addition of the step function results in an integral that is difficult to do explicitly.  
Instead, we begin with $M_1(0,0,0)=1$, and obtain the exact values of $M_\beta(n_1,n_2,n_3)$ for positive integer values of $\beta$ with the aid of recurrence relations. 
The moments for non-integer values of $\beta\geq 1$ are obtained by interpolating between adjacent integer-$\beta$ moments. 
The moments for $\beta$ between 0 and 1 are numerically computed by first performing numerical integration to obtain $M_0(n_1,n_2,n_3)$ and then interpolating with $M_1(n_1,n_2,n_3)$ to obtain the remaining non-integer $\beta$ values. 
The moment values for $\beta$ between $-\frac{1}{2}$ and $0$, because of the approach to the singularity at $\beta=-\frac{1}{2}$, require more care to compute numerically. 
However, since negative $\beta$ values turn out to not be needed except for values of $N(\lesssim 30)$ so small as to be irrelevant to tomography, we will focus only on computing the moments for $\beta\geq 0$. 
These steps are sufficient for us to assess the efficacy of the mean estimator, compute the minimax estimator over the allowed $\beta$ values, and compare the performance to other estimators.

The integral defining the moments can be simplified by using polar coordinates to parameterize the domain space, thereby getting rid of the delta function and the step function.
We write the probabilities $p_k$s in terms of the polar coordinates $r,\phi$ as in \eref{eq:pkrphi}. 
Without loss of generality, we can choose the trine orientation such that $\phi_0=0$. 
Then,
\begin{equation}
M_\beta(n_1,n_2,n_3)=\frac{2}{27}\int_0^1 \rmd r~r \int_{(2\pi)}\frac{\rmd \phi}{2\pi}~\prod_{k=1}^3{\left[1+r\cos(\phi-\phi_k)\right]}^{n_k+\beta-1}.
\end{equation}
The moments have permutation symmetry,
\begin{equation}
\fl M_\beta(k,l,m)=M_\beta(l,m,k)=M_\beta(m,k,l)=M_\beta(l,k,m)=M_\beta(m,l,k)=M_\beta(k,m,l),
\end{equation}
and obey a sum rule,
\begin{equation}
\fl M_\beta(n_1+1,n_2,n_3)+M_\beta(n_1,n_2+1,n_3)+M_\beta(n_1,n_2,n_3+1)=M_\beta(n_1,n_2,n_3),
\end{equation}
since $p_1+p_2+p_3=1$.
Note also a useful identity,
\begin{equation}
p_1p_2p_3={\left(\frac{1}{3}\right)}^3{\left[1-\frac{3}{4}r^2+\frac{1}{4}r^3\cos(3\phi)\right]}.
\end{equation}
\bigskip

\noindent\underline{Recurrence relations for integer values of $\beta\geq 1$}
\smallskip

We begin with the moment $M_1(0,0,0)=1$ and note that
\begin{equation}
M_\beta(n_1,n_2,n_3)=M_1(n_1+\beta-1,n_2+\beta-1,n_3+\beta-1).
\end{equation}
Hence, all moments for positive integer values of $\beta$ can be obtained from the moments $M_1(n_1,n_2,n_3)$ for $\beta=1$ with integers $n_1,n_2,n_3\geq 0$.
For notational simplicity, we drop the subscript `$1$' whenever we are discussing moments for $\beta=1$,
\begin{equation}
M(n_1,n_2,n_3)\equiv M_1(n_1,n_2,n_3)=2\int_0^1\rmd r~r\int_{(2\pi)}\frac{\rmd\phi}{2\pi}~p_1^{n_1}p_2^{n_2}p_3^{n_3}.
\end{equation}

The goal here is to develop recurrence relations that connect $M(n_1,n_2,n_3)$ to moments with smaller values of $n_k$s. Observe that differentiating the $p_k^{n_k}$ in the integrand of $M(n_1,n_2,n_3)$ will decrease its power from $n_k$ to $n_k-1$, taking a step towards our goal. To implement this differentiation, we make use of the two-dimensional gradient operator,
\begin{equation}
\vec{\nabla}\equiv \hat e_x\frac{\partial}{\partial x}+\hat e_z\frac{\partial}{\partial z}=\hat e_r\frac{\partial}{\partial r}+\hat e_\phi\frac{1}{r}\frac{\partial}{\partial \phi},
\end{equation}
and introduce the \emph{surface} moment, 
\begin{equation}
\fl\qquad L(n_1,n_2,n_3)\equiv \int_0^1\rmd r~r\int_{(2\pi)}\frac{\rmd\phi}{2\pi}~\vec\nabla\cdot{\left(\vec rp_1^{n_1}p_2^{n_2}p_3^{n_3}\right)}=\int_{(2\pi)}\frac{\rmd\phi}{2\pi}~p_1^{n_1}p_2^{n_2}p_3^{n_3}\Big\vert_{r=1},
\end{equation}
where $\vec r\equiv x\hat e_x+z\hat e_z=r\hat e_r$. In the second equality, we have used Gauss's theorem to convert the integral over the disk into an integral over its circumference ($r=1$). Performing the divergence operation in the integrand of $L$, we obtain the following relation,
\begin{eqnarray}
\fl\qquad &(N+2)M(n_1,n_2,n_3)-2L(n_1,n_2,n_3)\nonumber\\
\fl\qquad =&\frac{1}{3}{\Big[n_1M(n_1-1,n_2,n_3)+n_2M(n_1,n_2-1,n_3)+n_3M(n_1,n_2,n_3-1)\Big]}.\label{eq:MRecur}
\end{eqnarray}
The left side of the equation involves moments with total number of clicks $N=n_1+n_2+n_3$; the right side involves moments with $N-1$ clicks. Given the moments for $N-1$ clicks, we can use this recurrence relation to compute the moments for $N$ clicks, provided we can also compute the $L$ moments for $N$ clicks easily.

To obtain recurrence relations for the surface moments, we again try to differentiate the $p_k$s, this time with respect to $\phi$. Specifically, we begin with the identity
\begin{equation}
\int_{(2\pi)}\frac{\rmd\phi}{2\pi}~\frac{\rmd}{\rmd\phi}\sin\phi~p_1^{n_1}p_2^{n_2}p_3^{n_3}\Big\vert_{r=1}=0.
\end{equation}
Carrying out the $\phi$ derivative in the integrand above and re-expressing the result in terms of the $L$ moments, we obtain the relation
\begin{eqnarray}
\fl\quad &&3(N+1)L(n_1+1,n_2,n_3)\nonumber\\
\fl\quad&=&(N+1+n_1)L(n_1,n_2,n_3)+n_2L(n_1+1,n_2-1,n_3)+n_3L(n_1+1,n_2,n_3-1)\nonumber\\
\fl\quad &&-\frac{1}{2}n_2L(n_1,n_2-1,n_3)-\frac{1}{2}n_3L(n_1,n_2,n_3-1).\label{eq:LRecur}
\end{eqnarray}
The first line of the equation involves $N+1$ clicks, the second line involves $N$ clicks, and the last line $N-1$ clicks. Permutation symmetry yields similar recurrence relations for $L(n_1,n_2+1,n_3)$ and $L(n_1,n_2,n_3+1)$ in terms of moments involving fewer clicks.

The recurrence relations \eref{eq:MRecur} and \eref{eq:LRecur}, together with the initial values $M(0,0,0)=L(0,0,0)=1$, generate exactly all moments $M_\beta(n_1,n_2,n_3)$ for non-negative integers $n_1,n_2$ and $n_3$, and $\beta$ taking integer values $\geq 1$. One can now put these recurrence relations into a computer and efficiently compute, to a desired precision, the value of any moment with integer $\beta\geq 1$. For even better numerical accuracy and speed, one can also make use of additional formulas for special cases of the moments, such as
\begin{eqnarray}
M(n,0,0)&=2{\left(\frac{1}{3}\right)}^n\frac{n!}{(n+2)!}{\left[1+\sum_{k=1}^n{\left(\frac{1}{2}\right)}^k\frac{(2k)!}{(k!)^2}(k+1)\right]},\nonumber\\
L(n,0,0)&={\left(\frac{1}{6}\right)}^n\frac{(2n)!}{(n!)^2},\nonumber\\
L(n,1,0)&={\left(\frac{1}{6}\right)}^{n+1}\frac{(2n)!}{(n!)^2}~\frac{n+2}{n+1},\nonumber\\
L(n,2,0)&={\left(\frac{1}{6}\right)}^{n+2}\frac{(2n)!}{(n!)^2}{\left[1+\frac{2(4n+5)}{(n+1)(n+2)}\right]},\nonumber\\
L(n,n,n)&=2{\left(\frac{1}{216}\right)}^n\frac{(2n-1)!}{(n-1)!~n!}.
\end{eqnarray}
\bigskip

\noindent\underline{Moments for $\beta=0$ and non-integer $\beta$}
\smallskip

The moments $M_0(n_1,n_2,n_3)$ are computed approximately by direct numerical integration. 
One can verify the accuracy of the numerical integration routine by comparing the results to exact values like those for integer $\beta$, or to the exact value 
\begin{equation}
M_0(0,0,0)=\pi 12\sqrt{3},
\end{equation}
for which the integral can be done analytically. 

For non-integer values of $\beta> 0$, we interpolate between the two nearest integer $\beta$ values. 
Numerically, we find that the logarithm of $M_\beta$, for fixed $n_1,n_2,n_3$ values, is well-approximated by a linear function of $\beta$. 
An exponential interpolation, or equivalently, a linear interpolation of the logarithm of $M_\beta$, hence works well,
\begin{equation}\label{eq:Mexp}
\frac{M_\beta(n_1,n_2,n_3)}{M_{\lfloor\beta\rfloor}(n_1,n_2,n_3)}\approx {\left(\frac{M_{\lfloor\beta\rfloor+1}(n_1,n_2,n_3)}{M_{\lfloor\beta\rfloor}(n_1,n_2,n_3)}\right)}^{\beta-\lfloor\beta\rfloor},
\end{equation}
where $\lfloor\beta\rfloor$ denotes the largest integer $\leq\beta$.

For $n_1,n_2,n_3\sim N\gg 1$, we can understand this exponential behaviour.
When $n_k$s are large, the integrand of $M_\beta$, viewed as a function over the disk in Figure \ref{fig:triangleCircle}, is sharply peaked about the centre of the disk.
This means that it makes little difference whether we integrate only over physically allowed $p_k$s (those in the disk), or over all $p_k$s in the entire triangle.
This allows us to approximate $M_\beta(n_1,n_2,n_3)$ for the trine problem by that of the classical 3-sided die problem, or equivalently, to ignore the step function in $\chi(p)$.
Then, using the explicit formula for the classical moments given in \eref{eq:dieM} and invoking Stirling's formula $n!\approx \sqrt{2\pi}n^{n+1/2}e^{-n}$ to approximate the Gamma functions, we obtain
\begin{equation}
\frac{M_\beta(n_1,n_2,n_3)}{M_{\lfloor\beta\rfloor}(n_1,n_2,n_3)}\approx {\left(\frac{n_1}{N}\frac{n_2}{N}\frac{n_3}{N}\right)}^{\beta-\lfloor\beta\rfloor},
\end{equation}
exactly of the form of \eref{eq:Mexp}.
Numerically, we find that this exponential interpolation works well even for small $n_k$ values.

These recurrence relations and interpolations allow us to compute, as a ratio of moments (see \eref{eq:pkMERatio}), the mean estimator $\hat\rho_\mathrm{ME}^{(\beta)}$ for the trine for any $\beta\geq 0$ and any data $D_N\sim\{n_1,n_2,n_3\}$.
For integer $\beta\geq1$, the moments are exact, and the resulting estimated probabilities $(\hat p_k)_\mathrm{ME}^{(\beta)}$ are also exact.
For $\beta=0$, the estimated probabilities are accurate up to the precision of the numerical integration used to compute those moments.
For non-integer values of $\beta$ where interpolation is done, the moments, and likewise the estimated probabilities, are only approximate. 
In particular, because the exponential interpolation used for non-integer $\beta$ does not respect the constraint $\sum_k p_k=1$, we end up with estimated probabilities that violate this constraint, but only by a small amount (typically less than 3 percent). 
A simple remedy is to normalize the estimated probabilities, and we verify numerically that these normalized probabilities also satisfy the quantum constraint of $\hat p^2\leq \frac{1}{2}$.

\subsection{Minimax mean estimator for the trine}
\begin{figure}[ht]
\begin{center}
\includegraphics[width=0.75\textwidth]{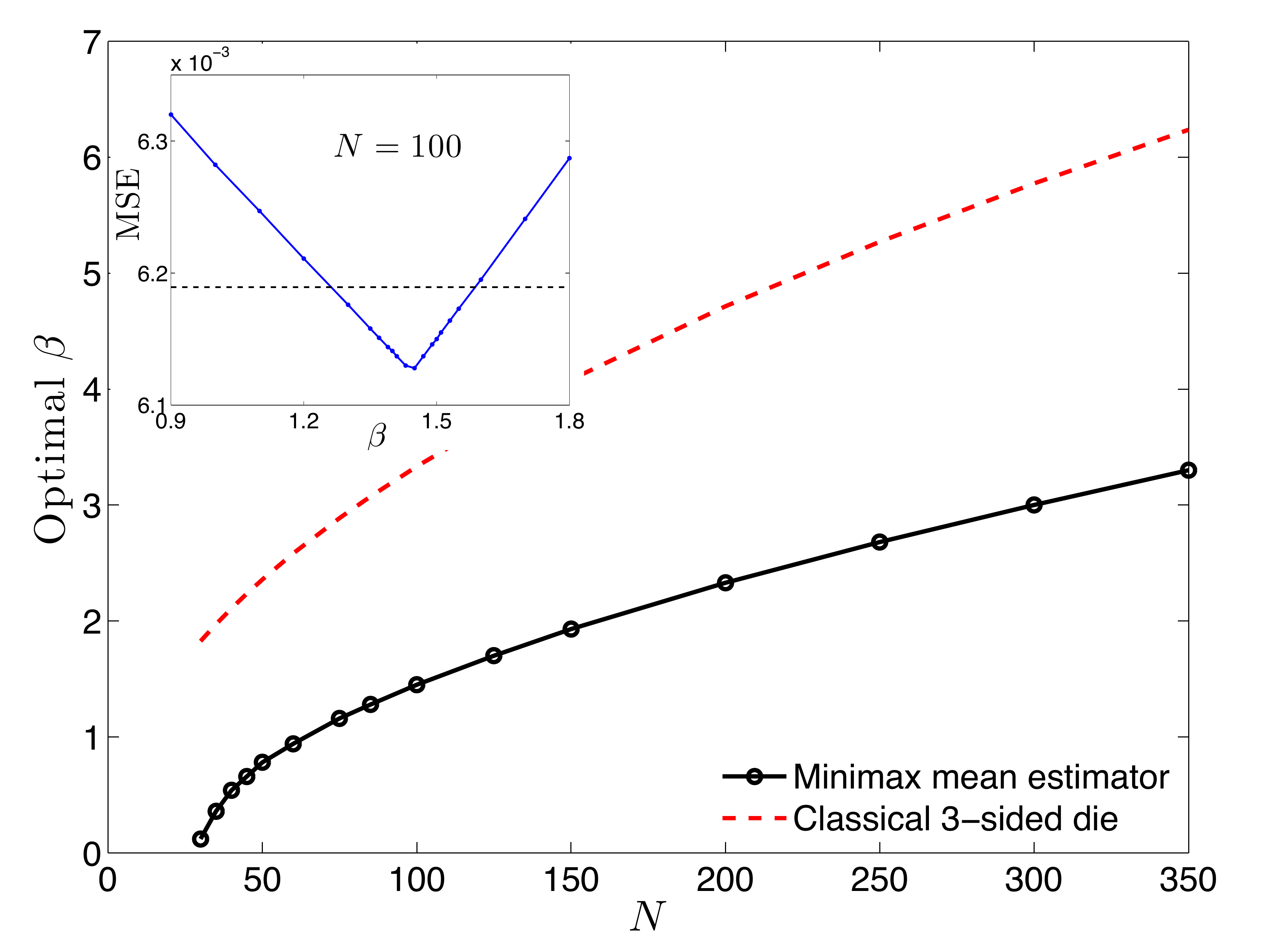}
\caption{\label{fig:beta} Plot of the optimal $\beta$ values for the minimax mean estimator applied to the trine problem versus $N$ (solid black line with circular markers). Shown on the same plot for comparison are the optimal $\beta$ values for the minimax estimator for the analogous 3-sided die problem (see \eref{eq:betaDie}) (dashed red line). Inset: The solid blue line gives the MSE as a function of $\beta$ for $N=100$; the dashed black line indicates the value of MSE that is one percent more than the minimum MSE value.}
\end{center}
\end{figure}

\begin{figure}[ht]
\begin{center}
\includegraphics[width=0.9\textwidth]{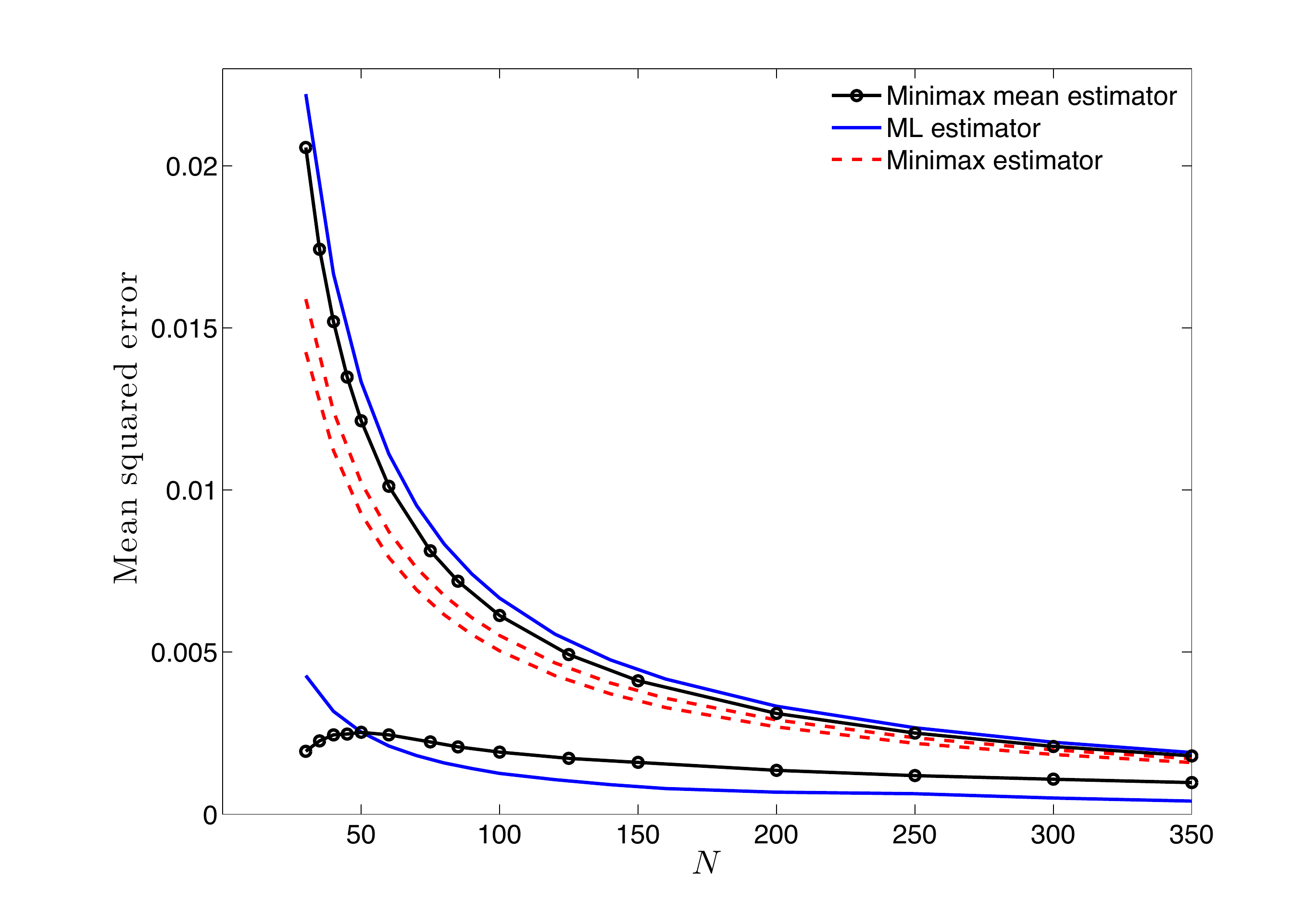}
\caption{\label{fig:MSEPlot} We plot the maximum and minimum MSE (over all states) for three different estimators: (1) the minimax mean estimator (using the optimal $\beta$ value for each $N$) (solid block line with circular markers), (2) the ML estimator for the trine measurement (solid blue line), and (3) the minimax estimator from \cite{Ng12} (dashed red line).}
\end{center}
\end{figure}

What remains is to choose the optimal value of $\beta$ to use.
As in the classical die problem, we choose $\beta$ by a minimax approach, and define the optimal $\beta$ as that which attains the minimax MSE,
\begin{equation}
\min_\beta~\max_\rho~ \textrm{MSE}{\left(\rho,\hat\rho_\mathrm{ME}^{(\beta)}\right)}.
\end{equation}
Unlike the classical die, we can only perform this optimization numerically. The optimal $\beta$ value, as well as the resulting minimax MSE, are reported in Figures \ref{fig:beta} and \ref{fig:MSEPlot}.

In Figure \ref{fig:beta}, we give the optimal $\beta$ value as a function of $N$, the total number of copies measured. 
The optimal $\beta$ starts close to zero for $N=30$ and increases monotonically as $N$ increases, in a manner suggestive of a $\sqrt{N}$ behaviour. 
In the same figure, we have also plotted the optimal $\beta$ value of $\sqrt{N}/3$ for the 3-sided die problem. 
The shapes of the two curves are qualitatively similar. 

One can understand qualitatively the offset in the $\beta$ values between the classical and the quantum problems. 
For the classical problem, for $\beta>1$, $f(p)$ is a function that is large near the centre of the triangle in Figure \ref{fig:triangleCircle}, and small near the boundary. 
The larger the value of $\beta$, the smaller the weight assigned to boundary states compared to states near the centre. For the trine problem,  $f(p)$ has a similar behaviour, but the states outside the disk are given zero weight because of the step function in $\chi$. 
This step function can be thought of, heuristically, as making the overall weight more peaked about the centre, as if $f(p)$ has a larger value of $\beta$. 
This reasoning agrees with the observation that the optimal $\beta$ for the trine problem is smaller than that for the same $N$ for the classical 3-sided die.

One should note that the minimum over $\beta$ for a given $N$ is not a sharp one. 
As illustrated in the inset of Figure \ref{fig:beta}, the value of the MSE for $N=100$ varies only by about one percent when we move about 0.2 away from the optimal $\beta$. 
This behaviour is also observed for other values of $N$. 
This means that, whether we take $\beta$ equal to the optimal value as given in Figure \ref{fig:beta}, or $\pm0.2$ of the optimal value, the performance of the mean estimator does not change significantly. 
Of course, if one moves too far from the optimal value of $\beta$, we see a marked increase in the MSE, but around the optimal point, the exact value of $\beta$ does not matter much.
After all, it is meaningless to compute the estimator to a precision beyond that justified by the data, which, in practice, will be polluted by some level of noise.

Figure \ref{fig:MSEPlot} gives the maximum and minimum MSE for the minimax mean estimator (with the optimal $\beta$ value for each $N$). 
Comparing these with the MSE for the ML estimator for the trine measurement (solid blue line) \cite{Rehacek04}, we see that the performance of our minimax mean estimator is slightly better in terms of slightly smaller maximum MSE, but the minimum error is slightly higher. 

We have also plotted the MSE for the corrected minimax estimator from \cite{Ng12} for the trine measurement. 
For this estimator, the maximum MSE is noticeably smaller than either the ML estimator or the minimax mean estimator. 
The minimum MSE is also significantly higher than either, but this actually gives the corrected minimax estimator the nice feature that the MSE is nearly constant over all states, as indicated by the close values of the maximum and minimum MSE. 
Given that we typically have little or no prior information about the state we are given, a constant MSE provides an objective way of treating every state in a fair manner.

The comparisons with the ML estimator and the corrected minimax estimator suggest that the minimax mean estimator discussed in this paper does not provide much advantage. 
If one is concerned with having a smaller minimum MSE (for example, if one knows that the states that attain this minimum value are more likely to occur), and does not mind the slightly higher maximum error, then, one would use the ML estimator. 
If one prefers a lower worst-case performance, or require a fair treatment of every state, one would use the corrected minimax estimator. 
The minimax mean estimator, despite its natural generalization from the minimax estimator for the analogous classical die problem, does not perform quite well enough to justify the complicated procedure required to compute it. 

An alternate way to generalize the minimax estimator for the classical die problem to the quantum case is to consider a different form of the weight function $f$. 
In our analysis above, because of the mathematical similarities with the classical die when viewed in terms of the probabilities $p_k$, it was natural to utilize exactly the same $f(p)$ as in the classical problem. 
Another possibility is to choose
\begin{equation}\label{eq:detRho}
f(p)={\left[\textrm{det}(\rho)\right]}^{\beta-1},
\end{equation}
where $\rho$ is considered as a function of the $p_k$s.
For the classical die, where every state $\rho$ can be expanded in terms of the basis that describes the measurement (see \eref{eq:dieState}), the determinant of $\rho$ is nothing but a product of the $p_ks$, giving the equivalence between the choice of $f$ either as in \eref{eq:f} or in \eref{eq:detRho}.
For the quantum case, the two choices differ.
\eref{eq:detRho} might also be a plausible choice of $f$ since it was previously used in \cite{BlumeKohout10a}, in the context of hedged maximum likelihood, as an additional weight function to generalize the same classical estimator in \eref{eq:addbeta} to the quantum regime.
With this alternate choice of $f$, one can employ the same techniques as described above to construct the minimax mean estimator for this weight function. 
Our preliminary investigations into this, however, indicate that the performance (in terms of the MSE) of the resulting estimator is very similar to that of the minimax mean estimator described before and the alternate choice of weight function offers no advantage over our previous choice.

In fact, considering an $f(p)$ that is a product of the $p_k$s, as we have done above, is more attractive than the choice in \eref{eq:detRho}. 
This is because the product of $p_k$s can be thought of as explicitly incorporating information about our choice of the tomographic measurement. 
This knowledge about the measurement used, from a perspective of interpreting the weight function as encompassing one's prior information, should enter the construction of the estimator.
On the other hand, $\textrm{det}(\rho)$ does not single out any particular measurement---one would write down the same function regardless of the measurement used---and it depends on the $p_k$s only implicitly through $\rho$.
In any case, either choice gives similar MSE values that do not quite outstrip the performance of previously known, simpler estimators.

\section{Conclusions}\label{sec:conc}

Motivated by the classical die problem, we derived the minimax mean estimator for the tomography of a qubit restricted to the $x$-$z$ plane of the Bloch sphere with the data acquired by a trine measurement. 
The trine problem has many similarities to the classical 3-sided die problem, with a single key difference being an additional physicality constraint imposed by quantum mechanics.
The similarities invite us to apply the same minimax approach used in the classical problem to look for a good estimator for the quantum problem.
That this would be a good idea is reinforced by our earlier work in \cite{Ng12} where a simple ad-hoc adaptation of the classical estimator to the quantum problem worked very well.
The mean estimator used for the classical problem also provides a very natural and elegant framework for incorporating the quantum constraint.
Nevertheless, we find, somewhat surprisingly, that the resulting minimax mean estimator does not offer much advantage over simpler estimators like the ML estimator or the corrected minimax estimator.
It yields slightly better worst-case performance than the ML estimator, but the small gain does not warrant the additional complications required to compute it.
This is a reminder to us how much more complex the quantum world can be, and how intuition from classical problems can sometimes fail in translation to the quantum case.

An important step forward will be to explore higher dimensions and other choices of tomographic measurements. 
As the dimension of the system grows, the numerical complexity will undoubtedly increase.
Nevertheless, it is pertinent for us to question if the above conclusions hold in higher dimensions, since the qubit is often a rather special case. 
Another possible future direction is to study the behaviour of the minimax mean estimator for a different figure-of-merit than the MSE. 
While the MSE is a convenient and often-used choice of estimation error, there are certainly scenarios for which alternative measures of assessing the performance of the estimation strategy (for example, the mean trace distance or relative entropy) will be more appropriate.

\ack 
This work is supported by the National Research Foundation and the Ministry of Education, Singapore.

\section*{References}

\bibliographystyle{unsrt}

\end{document}